# Low-frequency broadband acoustic metasurface absorbing panels

Jun Ji[1], Dongting Li[2], Yong Li[2] and Yun Jing[1,*]

[1]*Department of Mechanical and Aerospace Engineering, North Carolina State University, Raleigh, North Carolina 27695, USA*

[2]*Institute of Acoustics, School of Physics Science and Engineering, Tongji University, 200092 Shanghai, China*

* yjing2@ncsu.edu

## Abstract

A broadband sound absorption attained by a deep-subwavelength structure is of great interest to the noise control community especially for extremely low frequencies (20-100 Hz) in room acoustics. Coupling multiple different resonant unit cells has been an effective strategy to achieve a broadband sound absorption. In this paper, we report on an analytical, numerical and experimental study of a low-frequency broadband (50-63 Hz, one third octave band), high absorption (average absorption coefficient ≈93%), near-omnidirectional (0° - 75°) acoustic metasurface absorber composed of 4 coupled unit cells at a thickness of 15.4 cm (1/45 of the wavelength at 50 Hz). The absorption by such a deep-subwavelength structure occurs due to a strong coupling between unit cells, which is realized by carefully engineering geometric parameters of each unit cell, especially the judicious assignment of lateral size to each unit cell. To further broaden the bandwidth (50-100 Hz, one octave band), a design with 19 unit cells coupled in a supercell is analytically studied to achieve an average absorption coefficient of 85% for a wide angle range (0°-75°) at a thickness of 20 cm (1/34 of wavelength at 50 Hz). Two additional degrees of freedom, the lateral size of supercell and the number of unit cells in the supercell, are demonstrated to facilitate such a causally optimal design which is close to the ideally causal optimality. The proposed design methodology may solve the long-standing issue for low frequency absorption in room acoustics.

## I. Introduction

A sound absorber with a broadband and high absorption at a deep-subwavelength scale is of great interest in many occasions, such as room acoustics [1], automobiles and aerospace engineering [2]. A particular interest is to realize the so-called modal equalization (i.e., absorbing the normal mode frequencies of a room which usually fall below 100Hz) [3][4][5][6] for improving sound generation and speech interpretation. However, this is hindered by the inability of conventional sound absorbing materials to effectively remove low frequency sound [7].

The emergence of acoustic metamaterials [8][9] and acoustic metasurfaces [10][11] has enabled novel methods to design acoustic functional devices and has facilitated the development of new sound absorbing structures. To achieve the deep-subwavelength scale, one strategy is to use a very thin decorated membrane [12][13]. In such a design, however, a uniform and controlled tension of the membranes is needed, which leads to fabrication challenges and durability issues. Another strategy is to modify the geometry of the conventional Helmholtz resonator (HR) and the microperforated panel (MPP) [14] into space-coiling structures [15][16][17], embedded-neck structures [18][19] or multi-coiled structures [20]. Under the condition of impedance match or critical coupling [21][22], these designs can achieve a perfect sound absorption. Both the strategies above, however, have relatively narrow absorption bandwidth, which inevitably hinders practical applications. Some designs improve the bandwidth of single/identical resonator by tailoring the damping, such as increasing the intrinsic material damping [23] or utilizing a heavily overdamped condition [24]. Such designs, however, are either impractical or can hardly be applied to airborne sound absorption without a sacrifice of thickness [22].

To maximize the bandwidth of acoustic absorbers, a supercell with a number of different resonators [25][26][27][28][29][30][31][32][33][44] has been proposed as an effective design strategy. However, their corresponding thicknesses along the propagating direction are either thick or left with a potential for improvement. It is thus reasonable to ask: How much potential is there left for improvement? Or, for a target absorption spectrum, what is the minimum sample thickness required? These questions were addressed recently by [34] in which a target absorption is attained with the minimum sample thickness as dictated by the law of causality. However, the sponge, which is not desired in some harsh environmental conditions, is needed to achieve a causally optimal design for a broadband design, by remediating the negative absorption effects resulting from using a small, finite number of resonators. In addition, the examples demonstrated in that study pertain to broadband absorption at mid to high frequency ranges. To design a causally optimal sound absorber which can achieve a broadband, near-omnidirectional high absorption at extremely low frequencies is still a largely unsolved issue (though this is partially addressed by [20]).

To address the aforementioned issue, this paper provides a systematic procedure to design a metasurface sound absorbing panels which is composed of non-uniform unit cells with different embedded necks in a supercell. In the design procedure, three degrees of freedom (DOFs), which are the assignment of lateral sizes of unit cells, the number of unit cells and the lateral size of the supercell, are crucial factors to provide a broad range of surface impedance for a target absorption spectrum. Constrained by a desired absorption spectrum, a supercell at a minimum thickness is designed with the aid of optimization based on the Genetic algorithm (GA).

The paper is structured as follows: In section II, the metasurface geometry and the corresponding theoretical model are presented, along with the causal optimality recently introduced to sound absorption. In section III A, a design with a validation of simulation and experimental results is presented. To form the desired absorption spectrum over 50-63 Hz, a fairly weak coupling between unit cell is assumed and each unit cell is thus designed independently [25][46] to have a perfect absorption (an impedance-matching condition) at its specified resonant frequency. Section III B presents a design where the lateral sizes of unit cells are demonstrated to be instrumental to reduce the thickness by means of Genetic-Algorithm-based optimization. The thickness of the optimized structure is reduced by 23% without sacrificing the absorption performance. A strong coupling is observed in the optimized design and leads to a new avenue to designing a deep-subwavelength sound absorber. In section III C, a judicious tuning of a supercell's lateral size and the number of unit cells in a supercell, combined with the careful design of the lateral sizes of unit cells, is demonstrated to jointly facilitate a causally optimal [34] absorber covering 50-100 Hz. The causally optimal absorber is further compared with an ideal absorber at an ideally minimum thickness. Section IV concludes the paper.

## II. Theoretical model

### A. Description of metasurface geometry

Figure 1(a) shows the proposed metasurface with 2-dimensional periodic array of supercells. Figure 1(b) shows one supercell with different resonant units with embedded necks. The incident pressure is $p_{in} = P_{in} e^{-j(k_x x + k_y y - k_z z)}$, in which $k_x = k \sin\theta_{in} \cos\varphi_{in}$, $k_y = k \sin\theta_{in} \sin\varphi_{in}$ and $k_z = \cos\theta_{in}$. $\theta_{in}$ is the polar angle of incidence and $\varphi_{in}$ is the azimuthal angle of incidence. The time dependence $e^{j\omega t}$ is dropped for simplicity.

### B. Theoretical model for single resonant unit cell with embedded neck

As to the visco-thermal loss model for the sound absorption of a narrow neck, the Stinson's model [35][36], which works for a broad range of frequencies, is used. Therefore, we have,

$$Z_n = -\rho_0 c_0 \frac{2j \sin(\frac{k_c l_n}{2})}{\sqrt{(\gamma - (\gamma-1)\psi_h)\psi_v}} \tag{1}$$

where $\rho_0$, $c_0$ and $\gamma$ are density, speed of sound and the ratio of specific heats of air. $k_c$, $\psi_v$ and $\psi_h$ refer to the complex wave number, viscous function and thermal function of the embedded neck. All the geometric symbols are referred to Fig. 1(b). This model, which takes thermal loss into account for the long neck, has been well validated against experimental results recently [18]. Note that the conventional HR model [37] is only a special type of three main models in this generalized model, which is characterized by the relationship between the radius of the neck and the thickness of the viscous boundary layer [14][35][38].

Under a reasonable assumption that all dimensions of the resonant unit cells are much smaller than the working wavelength, the acoustic impedance for the irregular-shape cavity resulted from the embedded neck can be approximated as [18],

$$Z_c = -\frac{jS\rho_0 c_0^2}{\omega V} \tag{2}$$

where $\omega$ is the angular frequency. $S$ and $V$ are the lateral area and the volume of the cavity. $V = Sl_c - \pi\left(\frac{d}{2} + T_{\text{neck}}\right)^2 l$. It is reasonable to neglect the thermal and viscous loss here since all the dimensions of the cavity are much larger than the boundary layer thickness.

The overall impedance for the resonator is a sum of the impedances of the neck and cavity, which can be written as

$$Z = Z_n + Z_c + 2\sqrt{2\omega\rho_0\eta} + j\omega\rho_0\delta \qquad (3)$$

where $\eta$ is the dynamic viscosity of air. $2\sqrt{2\omega\rho_0\eta}$ is the resistance correction. $\delta = [1 + (1 - 1.25)\varepsilon] \times \left(\frac{4}{3\pi}\right)d$ is the mass end correction in which $\varepsilon = \frac{d}{\min(C_w, C_l)}$ is the ratio of the neck's diameter to the narrower side of the cavity.

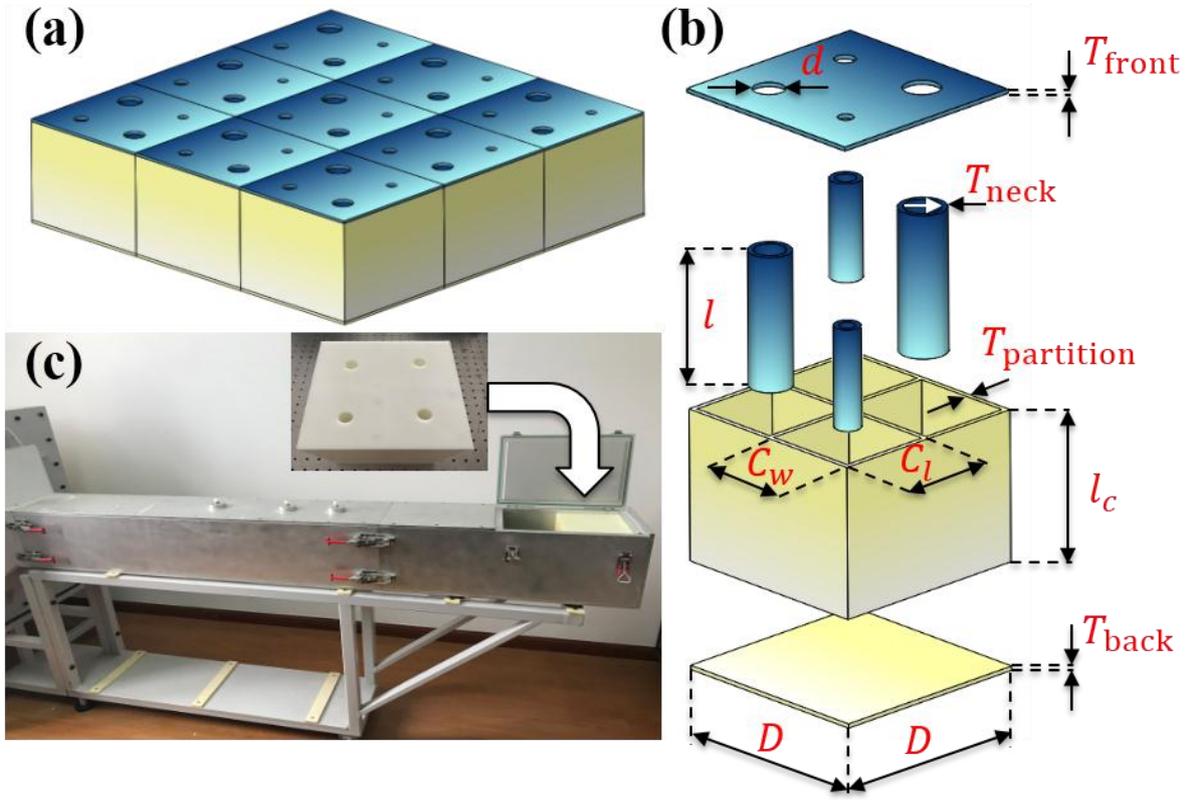

Figure 1. (a) Schematic of a metasurface absorber with 9 supercells. (b) The explosive view of one supercell with 4 different unit cells. The thickness of the front panel, $T_{\text{front}}$, the thicknesses of the necks' walls, $T_{\text{neck}}$, and the thicknesses of partitions, $T_{\text{partition}}$, are all set as 5 mm, according to our experimental trials, to mitigate the coupling effect between neighboring air domains so that the assumption of acoustically rigid boundaries in the analytical and numerical models is genuinely valid. The thickness of the back panel, $T_{\text{back}}$, is 2 mm. The diameter of the neck is $d$, the length of the neck is $l_n = l + T_{\text{front}}$, the lateral size of a unit cell is $S(C_w \times C_l)$, the length of the cavity is $l_c$, and the side length of a supercell is $D$. (c) The photograph of the experimental set-up for measuring absorption using the two microphones method. The lateral size of the impedance tube is 20 cm × 20 cm.

## C. Theoretical model for supercell with different unit cells

When a supercell of the panel is composed of $M$ different unit cells in parallel, its absorption performance can be characterized by a mean specific acoustic admittance or a mean specific acoustic impedance at normal incidence [43],

$$\frac{1}{Z_{tot}} = \sum_1^M \frac{\emptyset_N}{Z_N} \qquad (4)$$

Where $Z_N$ is the specific acoustic impedance for resonator $N$ with an end correction. $\emptyset_N$ is the surface porosity, $\emptyset_N = \pi \left(\frac{d_N}{2}\right)^2 / D^2$. The sound absorption coefficient can be subsequently calculated by [44],

$$\alpha(f, \theta_{in}) = \frac{4\text{Re}(Z_{tot})\rho_0 c_0 \cos\theta_{in}}{\{\rho_0 c_0 + \text{Re}(Z_{tot})\cos\theta_{in}\}^2 + \{\text{Im}(Z_{tot})\cos\theta_{in}\}^2} \qquad (5)$$

For normal incidence, a perfect absorption requires an impedance match between the panel and the air,

$$\text{Re}(Z_{tot}) = \rho_0 c_0 \qquad (6a)$$

$$\text{Im}(Z_{tot}) = 0 \qquad (6b)$$

Although this method is physically straightforward, computationally inexpensive and accurate for our deep-subwavelength structures, several underlying assumptions and limitations need to be noticed. First, Eq. 5 can be assumed to be independent [39] of the azimuthal angle $\varphi$ if the lateral size of the supercell is subwavelength. Otherwise, an asymmetric absorption [40][41] would be observed because of the geometric asymmetry, when the incident wave is from the same polar angle but different azimuthal angles. Second, this frequently used method is an approximation in terms of the accuracy of the mean specific impedance. Equation 4 is based on the assumptions [43] that lateral sizes of necks are considerably smaller than the wavelength and the lateral size of the supercell is sufficiently small, at which the pressure is constant over the surface and the volume displacement is additive. These assumptions can also be understood in a mathematical manner: Only the fundamental mode wave dominates the scattering field and thus needs to be considered, when $\lambda \geq 2D$ [44], regardless of the angle of incidence. When these assumptions are invalid, the couple-mode theory [40] and radiation impedance method [28][41] can be used, which either implicitly or explicitly take into account the coupling between unit cells.

Throughout this paper we will use a figure of merit $\alpha_{avg,\theta°}^{f_1 \sim f_2}$ to characterize and evaluate the absorption performance of the sound absorbing panel, which is the average absorption coefficient from $f_1$ to $f_2$ at the incident angle $\theta°$.

### D. Causal optimality in sound absorption

An absorber can be claimed to be causally optimal when its thickness reaches the minimum limitation dictated by the principle of causality [34]:

$$T \geq \frac{1}{4\pi^2} \frac{B_{eff}}{B_0} \left| \int_0^\infty \ln[1 - A(\lambda)] d\lambda \right| = T_{min} \qquad (7)$$

where $\lambda$ is the wavelength in air, $A(\lambda)$ is the absorption as a function of the wavelength, $B_{eff}$ is the effective bulk modulus of the absorber in the static limit, and $B_0$ is the bulk modulus of the air. $B_{eff}$ can be calculated as $\frac{B_0}{\phi_V}$, in which volume porosity $\phi_V \equiv V_{\text{air}}/V_{\text{tot}}$ is the volume fraction of the air domain considering $B_{\text{solid}} \gg B_{\text{air}}$.

**III. Results and discussion**

**A. Design for 50-63 Hz (one-third octave) absorption without optimization**

Motivated by the application of modal frequency absorption, four different resonant unit cells arranged in a supercell, with parameters listed in Table 1, are realized, with $\alpha_{avg,0°}^{50\sim63\text{Hz}} \approx 93\%$ for the analytical prediction and simulation, and $\alpha_{avg,0°}^{48\sim61\text{Hz}} \approx 93\%$ for the experiment, as shown in Fig. 2(a). The total thickness is 19 cm, which is around 1/36 of the wavelength at 50 Hz. Its absorption performance over 50-63 Hz as a function of the angle of incidence is also shown in Figs. 2(b)-(f). The analytical model is based on Eq. 5, while details of the sample fabrication and experimental setup are provided in Appendix A. Details of the numerical model can be found in Appendix B. The average absorption coefficient over 50-63 Hz is observed to be around 90% for angles of incidence smaller than 60 degree, because of the subwavelength size of the supercell. For angles of incidence larger than 60 degree, the absorption drops, likely due to an impedance mismatch. Interestingly, starting from around $50°$, only 3 absorption peaks appears instead of 4 peaks in the absorption spectrum, which is a signature of coupling between two unit cells [30] and this phenomenon is resulted from a coalescence of eigenstates [40][42] as the incident angle changes.

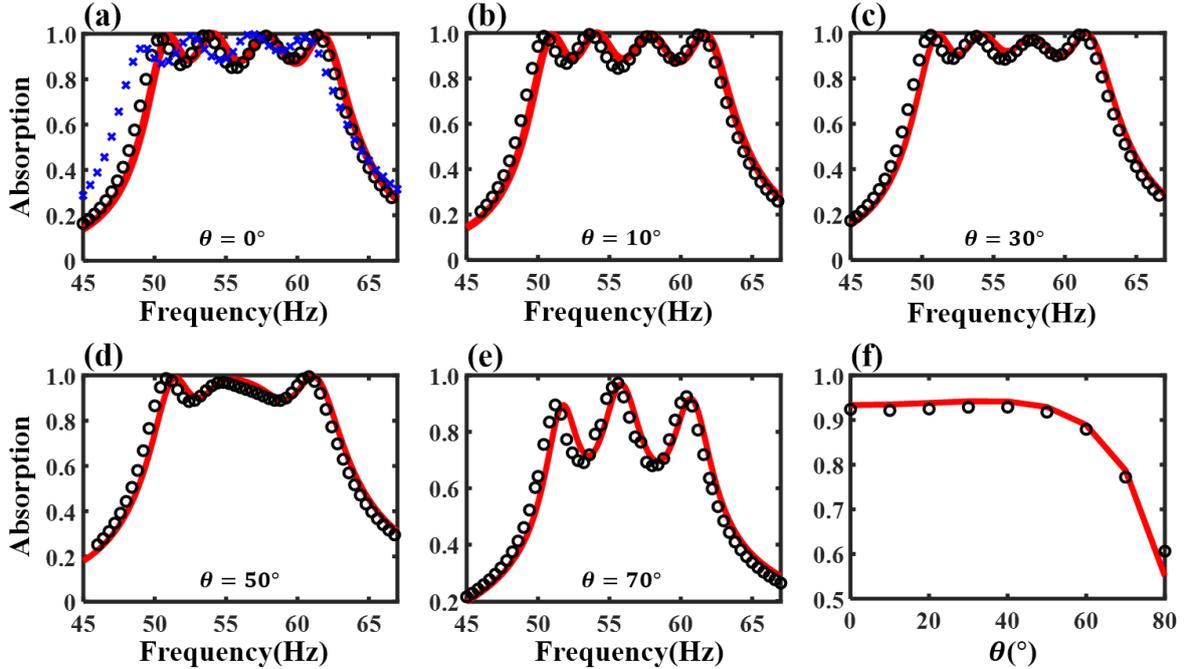

Figure 2. (a) The theoretical (red solid line), numerical (black circles) and experimental (blue crosses) absorption coefficient at normal incidence for the non-optimized metasurface absorber. (b)-(e) The theoretical and numerical

absorption coefficients for the angles of incidence 10°, 30°, 50° and 70°. (f) The theoretical and numerical average absorption coefficient over 50-63 Hz vs. angle of incidence.

| Number | $f$ (Hz) | $d$ (mm) | $l_n$ (mm) | $l_c$ (mm) | $S$ (mm$^2$) |
|---|---|---|---|---|---|
| 1 | 50.5 | 18.6 | 175.0 | | 102.0×93.0 |
| 2 | 54.0 | 18.2 | 165.0 | 183.0 | 92.0×92.0 |
| 3 | 58.1 | 17.4 | 123.0 | | 93.0×92.0 |
| 4 | 62.3 | 17.6 | 122.0 | | 83.0×93.0 |

Table 1. Geometric parameters of the metasurface absorber with 4 different unit cells without optimization. The lateral size of the supercell is 20 cm × 20 cm. The overall thickness of the absorber is 19 cm. Its volume porosity is 80%.

The design principle of this absorber is described as follows. The lateral size of the supercell is fixed as 20 cm × 20 cm, matching the size of our existing impedance tube (Fig. 1(c)). The suitable number of unit cells in a supercell, which depends on the $Q$ factor of unit cells and the desired absorption spectrum, is chosen as 4. Additional unit cells are needed, if their $Q$ factor is larger [20]. To form a continuously high absorption spectrum, each unit cell is designed to have a perfect absorption, following Eq.6, at a specific frequency inside the targeted absorption spectrum. The frequencies with perfect absorption are nearly equally-spaced. By assuming a fairly weak coupling effect between unit cells at these perfect absorption frequencies, each unit cell can thus be designed independently. During this design process, the lateral size of each unit cell is assigned without being optimized. In Fig. 3(a), the absorption coefficients of the supercells with only one individual unit cell are plotted as mark lines (blue diamonds, green circles, magenta points and black crosses), while that of the supercell with all unit cells is plotted as red solid line. For example, "unit cell 1" is a plot of $Z_{tot}(f) = Z_1(f)/\emptyset_1$, with $\emptyset_1 = \pi(d_1/2)^2/(20 \text{ cm})^2$. It is noted that the peaks of the red solid line, which indicates the near-perfect absorption of the supercell, matches well with peaks of the individual unit cells, which to a certain extent supports the assumption of the weak coupling between unit cells. However, it is still too early to claim from these analytical results in Fig. 3(a) that these near-perfect absorption points are contributed solely from individual unit cells operating independently. The underlying physical behavior will be revealed in the next paragraph by means of finite element method (FEM) simulations. Two important points of this trial-and-error design strategy should be mentioned. First, how large the cross-sectional area is assigned to each cavity is roughly proportional to the target wavelength because of the well-known concept that a larger volume of cavity is required for a smaller resonant frequency. For example, unit cell 1 which targets the lowest frequency has a cavity with the largest cross-sectional area. The best assignment of cross-sectional area to each cavity, however, is not investigated here but will be discussed in section B. Second, the design of a long and embedded neck plays a critical role in shortening the thickness of the unit cell for the longest wavelength and thus the overall thickness of the supercell, because of the signature of a large phase delay for a long neck [18]. During the design of unit cell 1, the neck length $l$ and the length of its cavity $l_c$ can be obtained by solving Eq. 6(a) and Eq. 6(b), when $d$ is parametrically swept and the cross-sectional area is set as 93 cm × 102 cm. One set of $l$ and $l_c$ is chosen, in which $l$ is as close as possible to $l_c$ but a clearance between $l$ and $l_c$ is left to ensure that the mass end correction is physically meaningful. The concept of curved necks [18], which is capable of further shrinking the total thickness, is not used here though, considering the fabrication complexity.

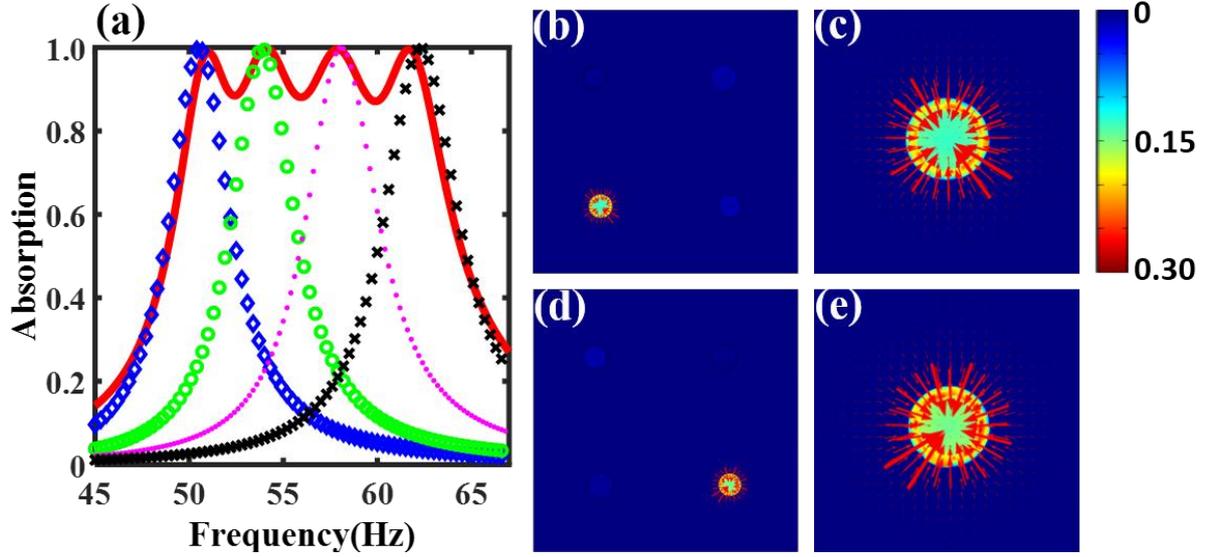

Figure 3. (a) The theoretical absorption coefficient of the non-optimized supercell (red solid line) and individual unit cells (blue diamonds, green circles, magenta points and black crosses are for unit cell 1,2,3 and 4 respectively). The curve for "unit cell $N$" is an absorption spectrum for a metasurface in which only unit cell $N$ can interact with the incident sound and all the other unit cells are blocked at the front panel. $N$=1,2,3,4. Sound intensity maps of the non-optimized supercell at (b) 53.6Hz and (d) 57.8Hz, respectively. The bottom-left unit cell of (b) and the bottom-right unit cell of (d) are enlarged, as shown in (c) and (d), respectively. The incident pressure is normalized to 1 Pa. Color bar (w/m$^2$): component of the sound intensity vector that are perpendicular to the supercell surface, i.e., $I_z$. Arrow: components of the sound intensity vector that are parallel to the supercell surface, i.e., $I_x$ and $I_y$. The size of the arrows is proportional to the magnitude of the intensity.

To gain insights concerning how the supercell is reacting at these near-perfect absorption points, the sound intensity fields just above the supercell surface are drawn at the resonant frequencies of two selected unit cells (53.6 Hz and 57.8 Hz) using COMSOL in Figs. 3(b)-(e), which provide another perspective to understand the underlying physics. At these resonant frequencies, the sound intensity is observed to be highly concentrated and drawn to the unit cell whose mode is excited. Instead of "penetrating" the entire supercell and being absorbed by all the unit cells at a similar amount, the sound energy is dissipated mostly by the excited unit cell while the other unit cells behave in a way as if they were "sealed". This is a clear signature of the weak coupling between the resonant unit cells at the resonant frequencies, and thus it is valid to design each unit cell independently for a near-perfect absorption.

**B. Design for 50-63 Hz (one-third octave) absorption with optimization**

In the previous section, the achieved design for the desired absorption spectrum has room to shrink the thickness, considering several deficiencies during the design methodology. First, by assuming a weak coupling between unit cells, all unit cells are designed independently to realize near-perfect absorption at their individual resonant frequencies to form a continuously high absorption spectrum. A better assignment of the contribution from all unit cells to the total absorption at each individual frequency of interest is left to be resolved. Second, the assignment of the cross-sectional area to each cavity follows an empirical idea: Generally, a larger cross-sectional area is required for a lower frequency if the cavities are required to have the length. A better assignment of cross-sectional areas also needs to be proposed. To address the two concerns above, the lateral areas of unit cells are assigned as additional variables to be treated in GA, while the thickness of absorber is to be optimized for the same absorption spectrum, i.e., $\alpha_{avg,0°}^{50\sim63\text{Hz}} \geq 93\%$. The details of variables, constraints and the object function are listed in Appendix C. Note that the area of supercell, instead of being a variable in the next section, is also fixed as 20cm by 20cm which is the size of our existing impedance tube.

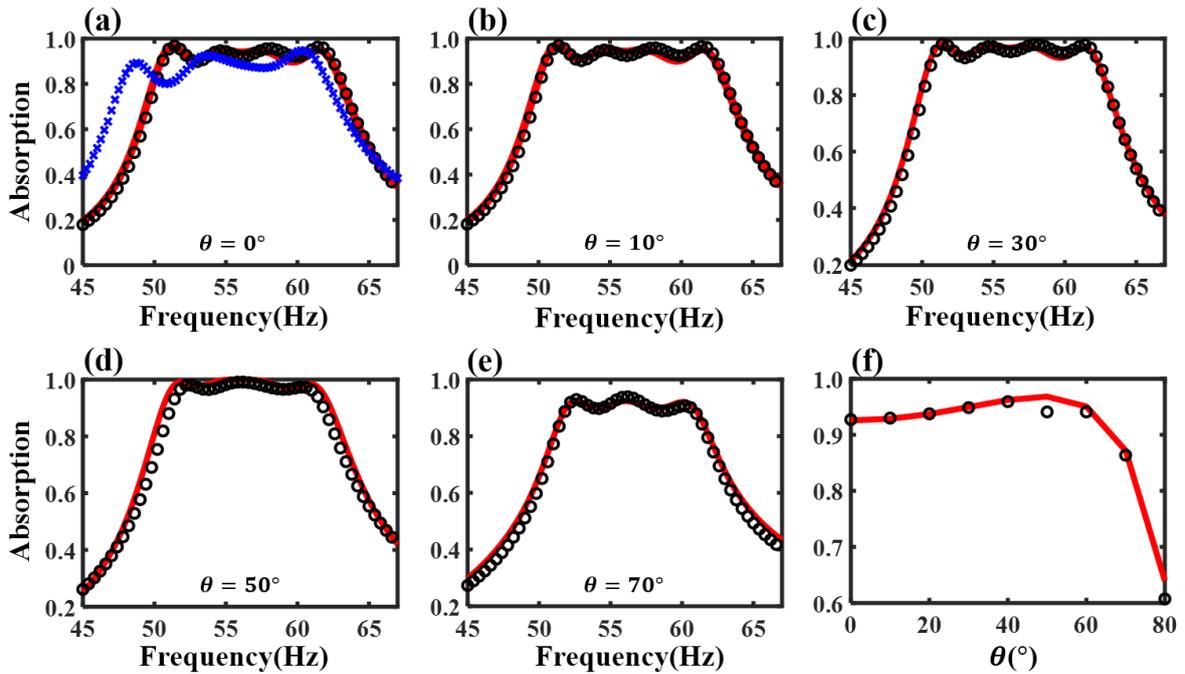

Figure 4. (a) The theoretical (red solid line), numerical (black circles) and experimental (blue cross) absorption coefficients at normal incidence for the optimized metasurface absorber. (b)-(e) The theoretical and numerical absorption coefficient for the angles of incidence 10°, 30°, 50° and 70°. (f) The theoretical and numerical average absorption coefficient over 50-63 Hz vs. angle of incidence.

| Number | $d$ (mm) | $l_n$ (mm) | $l_c$ (mm) | $S$ (mm$^2$) |
|---|---|---|---|---|
| 1 | 12.4 | 72.4 | | 112.1×99.9 |
| 2 | 9.9 | 63.2 | 147.0 | 80.7×88.1 |
| 3 | 7.6 | 21.1 | | 107.3×88.1 |
| 4 | 11.6 | 62.7 | | 75.9×99.9 |

Table 2. Geometric parameters of the optimized metasurface absorber composed of 4 different unit cells. The lateral size of the supercell is 20 cm × 20 cm. The overall thickness of the absorber is 15.40 cm. Its volume porosity is 84%.

Figure 4 shows the absorption performance of the optimized design, which is 15.4 cm (1/44.5 of the wavelength at 50 Hz) and is 23% thinner than the non-optimized one, with the geometric parameters listed in Table 2. However, its average absorption performance is not sacrificed as shown from the comparison of two designs' performance in Table 3. Note that the average frequency range for the analytical predictions and the numerical results is 50-63 Hz, while that for the experiment is 48-61 Hz because of the small frequency shift as shown in Fig. 2(a) and Fig. 4(a). The larger mismatch between the experimental result and the analytical prediction of the optimized design, compared with that of the non-optimized design, can be reasonably explained by a larger fabrication error, as shown in Appendix A. The maximum tolerance is 0.4 mm, which is larger than the previous one, which is 0.1 mm.

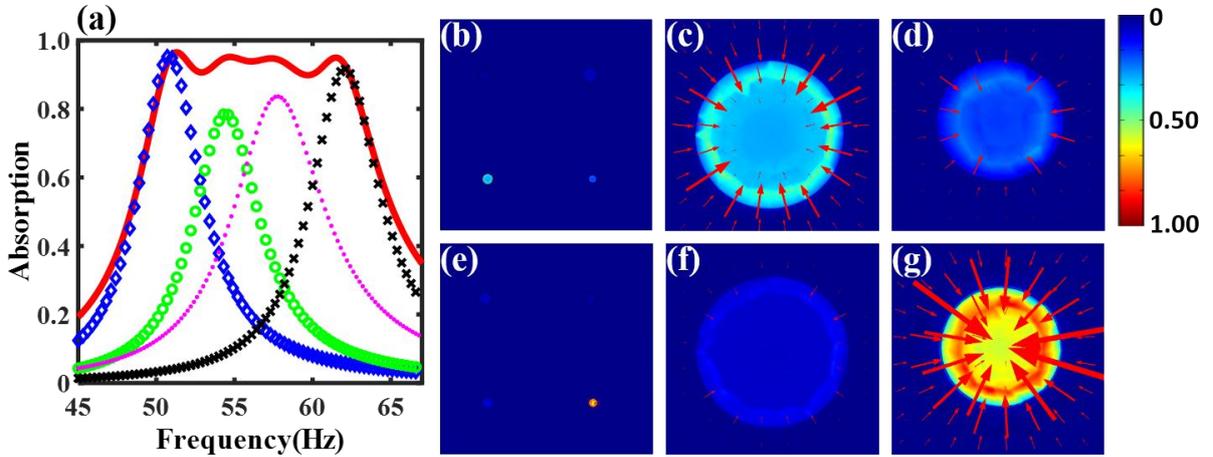

Figure 5. (a) The theoretical absorption coefficient of the optimized supercell (red solid line) and individual unit cells (blue diamonds, green circles, magenta points and black crosses are for unit cell 1,2,3 and 4 respectively). The curve for "unit cell $N$" is an absorption spectrum for a metasurface in which only unit cell $N$ can interact with the incident sound and all the other unit cells are blocked at the front panel. $N$=1,2,3,4. Sound intensity maps of the optimized supercell at (b) 54.4 Hz and (e) 57.8 Hz, respectively. The bottom-left unit cell and the bottom-right unit cell of (b) are enlarged, as shown in (c) and (d), respectively. The bottom-left unit cell and the bottom-right unit cell of (e) are enlarged, as shown in (f) and (g), respectively. The incident pressure is normalized to 1 Pa. Color bar (w/m$^2$): component of the sound intensity vector that are perpendicular to the supercell surface, i.e., $I_z$. Arrow: components of the sound intensity vector that are parallel to the supercell surface, i.e., $I_x$ and $I_y$. The size of the arrows is proportional to the magnitude of the intensity.

Next, we will uncover how the supercell thickness can be shrunk while a similarly good sound absorption performance is maintained. In contrast to the fairly weak coupling between unit cells demonstrated from the previous design at these near-perfect absorption points, a strong coupling, at which more than one unit cells jointly contribute to the total absorption, is observed as shown in Fig. 5(a). The intensity plots for the middle two local maxima of the absorption curve (i.e., 54.4 Hz and 57.8 Hz) are used to further shed light on the strong coupling. As shown in Figs. 5(b)-(d), two different unit cells are excited simultaneously and strongly coupled to work as energy sinks at 54.4 Hz. At 57.8 Hz, this phenomenon is not as pronounced as that at 54.4 Hz because the portion of absorption from one of the unit cells is large, as shown in Figs. 5(e)-(g). Interestingly, each individual unit cell is not required to satisfy the near-perfect absorption requirement [27] like shown in Fig. 3. Moreover, the assignment of cross-sectional areas to these non-uniform cavities is judiciously tuned thanks to the assistance of GA. As shown in Table 2, for the unit cell 1, which corresponds to the lowest resonant frequency, its cross-section area is 112.1 cm × 99.9 cm, which is 18% larger than that of the non-optimized design, i.e., 102.0 cm × 93.0 cm, while its neck and cavity lengths are shortened. Benefiting from the strong coupling and well-designed non-uniform cavities, the thickness of supercell, which is highly dependent on the neck and cavity lengths of unit cell with the smallest resonant frequency, is decreased from 190 mm to 154 mm by 23%. This is counter-intuitive to the long embedded and curled neck required [18] to achieve a large phase delay. However, this can be well explained from the perspective of the causality constraint (Eq. 7): The larger the volume porosity is, the smaller the minimum thickness is. Our optimized design, which has shorter embedded necks and thus a larger volume porosity, has a smaller actual thickness $T$ and a smaller causally minimum thickness $T_{\min}$ as well, as shown in Table 3.

|  | Theory | Simulation | Experiment | $\phi_V$ | $T$(mm) | $T_{\min}$(mm) |
|---|---|---|---|---|---|---|
| Design A | 93.3% | 92.4% | 93.1% | 0.80 | 190.0 | 169.7 |
| Design B | 92.6% | 92.7% | 87.8% | 0.84 | 154.0 | 149.1 |

Table 3. Comparison of the absorption performance between the non-optimized (design A) and the optimized (design B) absorber. "Theory" and "Simulation" refer to the average absorption coefficients over 50-63 Hz based on the analytical model and the numerical result, while "Experiment" refers to the average absorption coefficient over 48-61 Hz based on the experimental result. $\phi_V$ and $T$ are the volume porosity and total thickness of the design. $T_{\min}$ is the minimum thickness dictated by the principle of causality (i.e., Eq. 7).

## C. Design for 50-100 Hz (one octave) absorption with optimization

The aforementioned metasurface absorber, which benefits from the strong coupling between unit cells with judiciously designed cavities, is further explored to realize 50-100 Hz absorption. The lateral size of the supercell for an optimal design will be discussed below, instead of being assumed as 20 cm × 20 cm. For this reason, the experimental validation is not conducted in the section. The absorption performance of a resonator is conventionally [45] characterized by its maximum absorption cross section, which is in fact the lateral size of the supercell at which the perfect absorption condition (i.e., Eq. 6) is satisfied. The lateral size is also recently demonstrated to be directly related to the effective porosity [25][26] and thus the impedance matching condition in Eq. 6(a). Besides, the number of unit cells, which is a bridge between the absorption ability (e.g., bandwidth) of a single unit cell and the overall absorption of the supercell, is an additional important variable not well discussed in previous literatures.

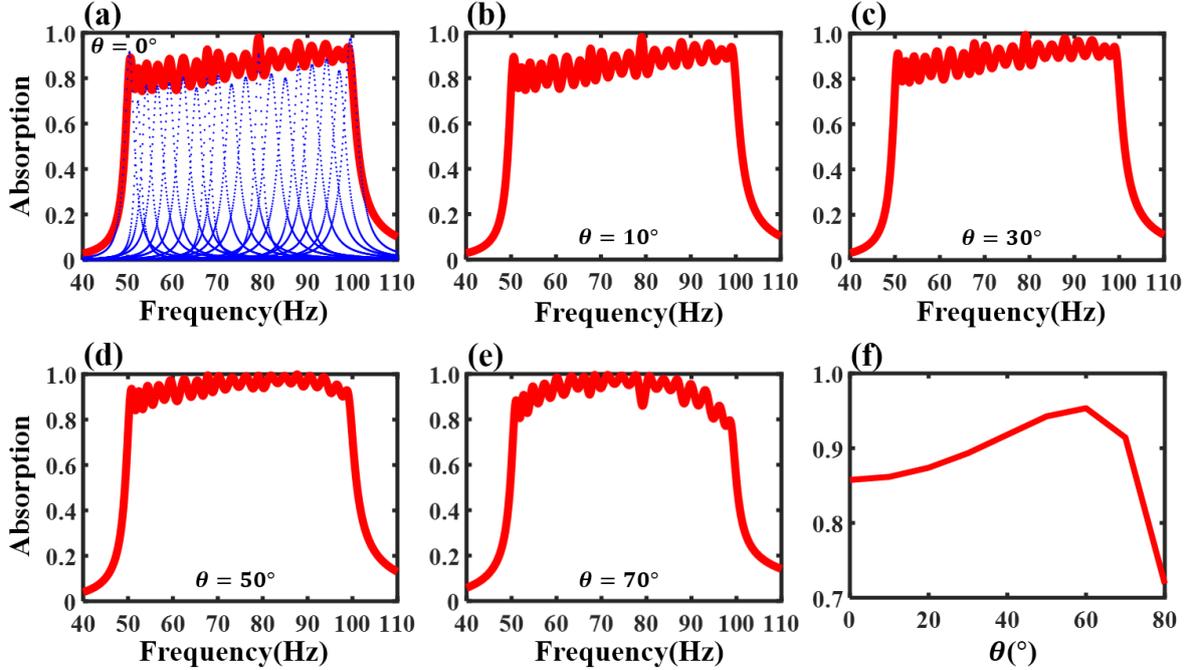

Figure 6. (a) The absorption coefficient of the optimized supercell (red solid line) and individual unit cells (blue points lines) at normal incidence. (b)-(e) The theoretical absorption coefficients for the angles of incidence 10°, 30°, 50° and 70°. (f) The theoretical average absorption coefficient over 50-100 Hz vs. angle of incidence.

With the above concerns, GA is used as an efficient tool to seek an optimal design which satisfies $\alpha_{avg,0°}^{50\sim100\text{Hz}} \geq 85\%$, in which the two additional variables, (the lateral size of supercell $D$ and the number of unit cells $M$), are incorporated. The details of variables, constraints and the object function are listed in Appendix D. An optimized design, at a thickness of 20 cm (which is 1/34.3 of the wavelength at 50 Hz) and a lateral size of 78.84 cm × 78.84 cm, is achieved with all the other parameters listed in Table 4. Within the supercell, there are 19 different unit cells whose lateral sizes are well tuned. All the unit cells for these three designs belong to the HR model regime according to the relationship between radii and thickness of viscous boundary layer [14][35]. Figure 6(a) shows the analytically predicted absorption for the supercell and individual unit cells. The analytical model is valid since the dimensions of the supercell satisfy the assumptions discussed in Section II C. The FEM validation is not provided here since the existing computation resource in our lab is insufficient for this design with 19 unit cells. The analytical prediction, however, has been proven to be accurate in previous two designs. The average absorption coefficient over 50-100 Hz as a function of angle of incidence is plotted in Fig. 6(f) and remains above 85% until 75 degree, which again demonstrates its near-omnidirectional absorption capability. The absorptions at 10,30,50 and 70 degree are plotted in Figs. 6(c)-(e), respectively.

| Number | $d$ (mm) | $l_n$ (mm) | $l_c$ (mm) | $S$ (mm$^2$) |
|---|---|---|---|---|
| 1 | 38.4 | 122.9 | | 145.6×137.4 |
| 2 | 21.1 | 11.9 | | 145.6×137.4 |
| 3 | 34.4 | 138.3 | | 145.6×137.4 |
| 4 | 26.1 | 33.1 | | 145.9×137.4 |
| 5 | 30.4 | 145.7 | | 180.8×137.4 |
| 6 | 20.1 | 5.6 | | 144.8×176.0 |
| 7 | 19.2 | 5.3 | | 147.8×176.0 |
| 8 | 17.5 | 5.2 | | 149.3×176.0 |
| 9 | 25.6 | 8.9 | | 157.4×176.0 |
| 10 | 17.7 | 5.0 | 193.0 | 164.2×176.0 |
| 11 | 20.8 | 17.9 | | 144.1×211.6 |
| 12 | 17.1 | 7.1 | | 146.9×211.6 |
| 13 | 19.1 | 16.8 | | 149.5×211.6 |
| 14 | 17.1 | 5.0 | | 150.4×211.6 |
| 15 | 22.6 | 39.1 | | 172.5×211.6 |
| 16 | 19.5 | 18.5 | | 156.3×243.4 |
| 17 | 17.3 | 7.9 | | 172.3×243.4 |
| 18 | 17.2 | 10.8 | | 180.7×243.4 |
| 19 | 23.3 | 22.5 | | 254.1×243.4 |

Table 4. Geometric parameters of an optimized metasurface absorber composed of 19 different unit cells. The lateral size of the supercell is 78.84 cm × 78.84 cm. The overall thickness of the absorber is 20 cm. Its volume porosity is 90%.

It is noted that the achieved design is causally optimal: If we insert the achieved absorption spectrum (the red curve in Fig. 7) of our design and the corresponding volume porosity (90% for this design) into Eq. 7, $d_{min}$ is calculated as 20 cm, which is precisely the thickness of our design.

Next, we will uncover how the two additional design variables, like the geometric parameters of all unit cells, play critical roles in reaching the causally optimal design (the minimum thickness) for a desired absorption spectrum. For the lateral size of supercell being a suitable value, on one hand, it should be as large as possible so that the lateral size of unit cells can be larger and therefore the thickness of all unit cells can be smaller to meet the design objective. On the other hand, it should not become too large because that will result in an impedance-mismatch for individual unit cells and thus a drop of absorption coefficient. The number of unit cells should also be properly chosen. If the number of unit cells is too small, it would be insufficient to achieve such a broad absorption spectrum by coupling.

However, if the number of unit cells is too large and the thickness is preserved, it would call for an increased lateral size of the supercell, which would result in a reduction of the impedance match and the absorption performance. Similarly, for such a case of a large number of unit cells, if the lateral size of the supercell is chosen to be preserved for the desired absorption performance, it would result in an increase in the thickness of the supercell.

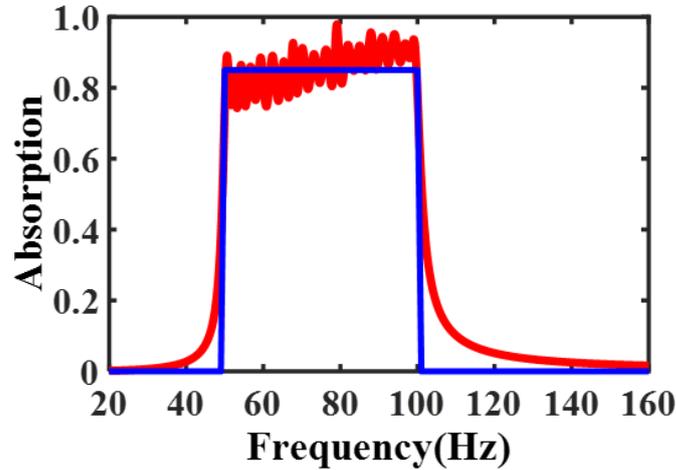

Figure 7. The absorption coefficient of the optimized supercell (red solid line) and the ideal absorber (blue solid line) at normal incidence. The optimal thickness for the optimized design is 20 cm. The optimal thickness for the ideal design is 16.9 cm when the volume porosity is set as 1 and the achieved absorption spectrum exactly matches with the desired one (85% over 50-100 Hz) in Eq. 7.

In addition, we would like to further comment on the impact of the causality constraint on guiding the design of the thinnest absorber for a desired absorption spectrum. First, it is not difficult to design an absorber which is causally optimal as indicated in the supplemental material of reference [34]: The MPP or sponge can easily become causally optimal if the desired absorption is not required to be high over a broad low frequency range. However, it is hard to realize a high absorption at such a low-frequency and broad range as presented in the paper, i.e., 50-100 Hz. Second, as shown in Eq. 7, the optimal thickness not only relies on the achieved absorption spectrum, but also is highly dependent on the volume porosity. The larger the volume porosity is, the smaller the optimal thickness is. In other words, a better space utilization makes an absorber thinner. A higher volume porosity ($\geq 95\%$) can be achieved by making the partitions between the unit cells as thin as 1 mm, at which the partitions may not be treated as ideally rigid anymore. Third, designing an absorber whose absorption spectrum is as close as possible to the desired one is crucial to shrink the minimum thickness, especially for low frequencies. As shown in Fig. 7, the absorption spectrum of an ideal absorber, whose absorption spectrum exactly matches with the desired one (85% over 50-100 Hz in this case), is plotted as a benchmark. The proposed design exhibits an absorption performance fairly close to the desired one with little unwanted absorption outside the frequency range of interest. This is due to the fine manipulation of acoustic surface impedance and the absorption spectrum by utilizing unit cells with large Q factors.

The optimal thickness of the ideal absorber is 16.9 cm by substituting a volume porosity of 100% and the desired absorption spectrum (the blue curve in Fig. 7) into Eq. 7. Thus, our 20 cm design is not only causally optimal, but is also close to the ideal design. This is benefited from an adequate tuning range of acoustic impedances, which is provided by the three DOFs during the design of supercell, to simultaneously shape the absorption spectrum and to maintain a high volume porosity. The three DOFs in fact provide a similar functionality as the sponge does in [34] where lateral sizes of unit cells are uniform, and the lateral size of the supercell as well as the number of unit cells are fixed.

**IV. Conclusion**

This paper analytically, numerically and experimentally studies a low-frequency broadband (50-63 Hz, one third octave band), high absorption (average absorption coefficient ≈ 93%), near-omnidirectional ($0°$ - $75°$) acoustic metasurface absorber composed of 4 coupled unit cells at a thickness of only 15.4 cm (1/45 of the wavelength at 50 Hz), in which the suitable assignment of lateral size to each unit cell plays a critical role in reducing the overall thickness. Furthermore, to realize a broadband and near-perfect absorption at an optimal thickness, the lateral size of supercell and the number of unit cells in the supercell are also judiciously engineered. Taking all these new degrees of freedom into account and with the assistance of Genetic Algorithm, a causally optimal acoustic absorber, which is a broadband (50-100 Hz, one octave band), high-absorption (average absorption coefficient 85%), near-omnidirectional ($0°$- $75°$) and deep-subwavelength (1/34 of wavelength at 50 Hz), is proposed. Our future work is to experimentally validate the proposed design for 50-100 Hz and further push the design of absorbers to the ideal one by means discussed in this paper. We believe that our design philosophy, in which three degrees of freedom are proven to be critical in reaching a causally optimal thickness, would offer a gateway to designing broadband sound absorbers with different formats of unit cells (MPP, HRs, space-coiling structures, etc.) and to meet the long-standing need in room acoustics.

**Acknowledgments**

Y. L. acknowledges support from the National Natural Science Foundation of China under Grant No. 11704284.

J.J. and D.L. contributed equally to this work.

**Appendix A: Sample fabrication and experimental setup**

The sample is fabricated by using 3D-printing involving laser sintering stereo-lithography (SLA) and photosensitive resin (UV curable). The nominal precision is 0.1mm. For design A, the measured diameters of necks are 18.50 mm, 18.17 mm, 17.38 mm and 17.60 mm, while the desired ones are 18.6 mm, 18.2 mm, 17.4 mm and 17.6 mm, as shown in Table 1. Thus, the maximum tolerance is 0.1 mm. For design B, the measured diameters of necks are 12.22 mm, 9.50 mm, 7.25 mm and 11.35 mm, while the desired ones are 12.4 mm, 9.9 mm, 7.6 mm and 11.6 mm, as shown in Table 2. Thus, the maximum tolerance for design B is 0.4 mm. The inconsistent tolerance might be a result of the quality of the resin.

The absorption coefficient is tested with a self-made impedance tube whose cross section is 20 cm by 20 cm, as shown in Fig. 1(c). The measurements of the absorption coefficient were performed using the impedance tube method complying with ASTM C384-04(2011) and ASTM E1050-12. Two 1/4-inch condenser microphones (Brüel & Kjær type-4187) are situated at designated positions to obtain the amplitude and phase of the pressure distribution. A digital signal (white noise) generated by the computer was sent to the power amplifier (Brüel & Kjær type-2734) and then powered the loudspeaker. By analyzing the signals from the two microphones, the absorption coefficient can be obtained.

**Appendix B: Numerical simulation**

The numerical simulation is performed using the Acoustic-Thermoacoustic Module of COMSOL Multiphysics 5.4. Owing to the large difference between the acoustic characteristic impedances for solid and air, the walls of the necks, the front panel, the back panel and the partitions are assumed to be sound-hard boundaries in all simulations, with no-slip and isothermal boundary condition applied. The numerical absorption coefficient is calculated by a ratio of the total dissipated power to the incident power. The total dissipated power is acquired by a volumetric integral of a default variable called total thermo-viscous power dissipation density in COMSOL. To ensure the accuracy of simulation, a mesh independent test is carried out and solutions converge when the largest element size is set as 1/6 of the smallest working wavelength, the largest element size at the neck region is refined as 1/4 of the corresponding radius, the number of boundary layer is set as 6 and the thickness of first layer is set as 1/5 of viscous boundary layer thickness. A Floquet periodicity is applied to the four side boundaries of the supercell.

**Appendix C: GA optimization for 50-63Hz (one-third octave)**

Genetic Algorithm in the toolbox of MATLAB is used to optimize our designs. The population size is 1000 and the crossover fraction is 0.8. Elite count, which selects the fittest individuals to dominate the population, is set as 20. The optimization stops after 50 stall generations. Due to the nature of heuristics, the algorithm is run 30 times for each design to find the globally optimal result. Parallel computing in MATLAB is used to minimize the computing time. To further speed up the optimization, the following design variables and constraints should be normalized.

For the design aimed at $\alpha_{avg,0°}^{50\sim63\text{Hz}} \geq 93\%$, there are 16 design variables:

$$X = (d_1, d_2, d_3, d_4, l_1, l_2, l_3, l_4, l_{c,1}, l_{c,2}, l_{c,3}, l_{c,4}, S_1, S_2, S_3, S_4)$$

The object is to minimize $F = \max(l_{c,1}, l_{c,2}, l_{c,3}, l_{c,4})$, which subjects to:

$$\begin{cases} \alpha_{avg,0°}^{50\sim63\text{Hz}} \geq 93\% \\ 0.1 \leq \dfrac{d_N/2}{\sqrt{\dfrac{\eta}{2\pi\rho_0 f_2}}} \leq 100, \quad \text{for } N = 1,2,3,4 \\ T_{\text{front}} \leq l_N \leq \dfrac{1}{4}\dfrac{c_0}{f_1}, \quad \text{for } N = 1,2,3,4 \\ 0 \leq S_N \leq D^2, \quad \text{for } N = 1,2,3,4 \\ l_N \leq l_{c,N} \leq \dfrac{1}{4}\dfrac{c_0}{f_1}, \quad \text{for } N = 1,2,3,4 \\ 4 \times \left(\dfrac{d_N}{2} + T_{\text{neck}}\right)^2 \leq S_N, \quad \text{for } N = 1,2,3,4 \\ \sum_{N=1}^{4} S_N + S_{\text{partition}} \leq D^2 \end{cases}$$

For the constraints above, the first one is the desired absorption performance. All the other constraints are geometric constraints, which are used to narrow the design space. The second constraint is the relationship between the radii of necks and the viscous boundary layer. The third and the fifth constraints require the lengths of the necks and cavities to be smaller than one quarter of the longest working wavelength. The fourth, the sixth and the seventh constraints require the lateral sizes of the cavities, the necks and the supercell to be feasible for manufacturing.

**Appendix D: GA optimization for 50-100Hz (one octave)**

For the Design aimed at $\alpha_{avg,0°}^{50\sim100\text{Hz}} \geq 85\%$, there are 4$M$+2 ($M$ is defined in section II C as the number of unit cells in a supercell) design variables:

$$X = (D, d_1, d_2, d_3, \ldots, d_M, l_1, l_2, l_3, \ldots, l_M, l_{c,1}, l_{c,2}, l_{c,3}, \ldots, l_{c,M}, S_1, S_2, S_3, \ldots, S_M)$$

The object is to minimize $F = \max(l_{c,1}, l_{c,2}, l_{c,3}, \ldots, l_{c,M})$, which subjects to:

$$\begin{cases} \alpha_{avg,0°}^{50\sim100\text{Hz}} \geq 85\% \\ 0.2 \leq D \leq \dfrac{1}{2}\dfrac{c_0}{f_1} \\ 0.1 \leq \dfrac{d_N/2}{\sqrt{\dfrac{\eta}{2\pi\rho_0 f_2}}} \leq 100, \quad \text{for } N = 1,2,3,\ldots,M \\ T_{\text{front}} \leq l_N \leq \dfrac{1}{4}\dfrac{c_0}{f_1}, \quad \text{for } N = 1,2,3,\ldots,M \\ 0 \leq S_N \leq D^2, \quad \text{for } N = 1,2,3,\ldots,M \\ l_N \leq l_{c,N} \leq \dfrac{1}{4}\dfrac{c_0}{f_1}, \quad \text{for } N = 1,2,3,\ldots,M \\ 4 \times \left(\dfrac{d_N}{2} + T_{\text{neck}}\right)^2 \leq S_N, \quad \text{for } N = 1,2,3,\ldots,M \\ \sum_{N=1}^{M} S_N + S_{\text{partition}} \leq D^2 \end{cases}$$

For the constraints above, the first one is the desired absorption performance. All the other constraints are geometric constraints, which are used to narrow the design space, as discussed in Appendix C.

By parametrically sweeping $M$ from 15 to 25, the overall thickness is found to reach the causally minimum thickness when $M = 19$.